# Application of Principal Component Analysis in Chinese Sovereign Bond Market and Principal Component-Based Fixed Income Immunization

Lim Tze Yee, Tony She, Kezia Irene
**School of Economics**
**Peking University**

**Abstract:**

This paper analyses the Chinese Sovereign bond yield to find out the principal factors affecting the term structure of interest rate changes. We apply Principal Component Analysis (PCA) on our data consisting of Chinese Sovereign bond since January 2002 till May 2018 with different yield to maturity. Then we will discuss multi-factor immunization model (method on hedging market risk) on bond portfolio.

***Key words:*** Principal Component Analysis(PCA), Slope, Level, Curvature, VaR Analysis, Multi-factor Immunization model

## Introduction

In our paper, we are using Principal Component Analysis (PCA), a dimension-reduction tool to revealed three main principal components that are sufficient in explaining the variation in the interest rate changes of China Sovereign Bond Market. PCA is proved to be an effective method of using the historical spot rates changed data to determine the best single direction and the best set of fundamental directions in which to anticipate spot rate changes. PCA will provide us three significant factors which is level, slope and convexity of the security. Based on the result, principal component durations and convexity can be computed from the first and second partial derivatives of the security with respect to the three factors of PCA. As we known, duration is the most commonly used risk measure for measuring the risk exposure of a security, while convexity usually complements duration to provide a closer approximation to interest rate risk. Therefore, calculate duration and convexity based on principal component provide us a more significant results. Furthermore, in our experiment, VaR analysis is applied by using principal component model in order to provide an accurate description of interest rate dynamics while maintaining a low number of risk factors. Lastly, according to the result we make a conclusion that investor can hedge the market risk of fixed income security portfolio by keeping the three factors, level, slope and convexity in a neutral condition. A multi-factor immunization model will be discussed in creating an immunized portfolio.

## Introduction of Principal Component Analysis (PCA) and Principal Components

Principal Component Analysis (PCA) is a dimension-reduction or data compression tool that used to reduce a large set of possibly correlated variables into a smaller set number of uncorrelated variables called principal components, that contains most of the information in the large set. By construction, the first principal component explains the maximum percentage of the total variance of interest change. The second is linearly independent from the first, and explain the maximum percentage of the remaining variance, and so on. The first principal component accounts for as much of the variability in the data as possible, and each succeeding component accounts for as much of the remaining variability as possible.

Since the PCA model explicitly selects the factors based upon their contributions to the total variance of interest rate changes, it may help in hedging efficiency when using only a small number of risk measures. Factor analysis is a general name denoting a class of procedures primarily used for data reduction and summarization. Factor analysis is an interdependence technique in that an entire set of interdependent relationships is examined without making the distinction between dependent and independent variables. Factor analysis is used in the following circumstances: identify underlying dimensions, or factors, that explain the correlations among set of variables; identify a new, smaller, set of uncorrelated variable to replace the original set of correlated variables in subsequent multivariate analysis (regression or discriminant analysis); To identify a smaller set of salient variables from a larger set for use in subsequent multivariate analysis.

Mathematically, each variable is expressed as a linear combination of underlying factors. The covariation among the variables is described in terms of a small number of common factors plus a unique factor for each variable. If the variables are standardized, the factor model may be represented as:

$$X_i = A_{i1}F_1 + A_{i2}F_2 + A_{i3}F_3 + \cdots + A_{im}F_m + V_iU_i$$

Where

$X_i$ = i th standardized variable

$A_{ij}$ = standardized multile regression coefficient of variable I on common factor j

$F_i$ = common factor

$V_i$ = standardized regression coefficient of variable I on unique factor i

$U_i$ = the unique factor of variable i

m = number of common factors

The unique factors are uncorrelated with each other and with the common factors. The common factors themselves can be expressed as linear combinations of the observed variables.

$$F_i = W_{i1}X_1 + W_{i2}X_2 + W_{i3}X_3 + \cdots + W_{ik}X_m$$

Where

$F_i$ = estimate of i th factor

$W_i$ = weight or factor score coefficient

k = number of variables

It is possible to select weights or factor score coefficients so that the first factor explains the largest portion of the total variance. Then a second set of weights can be selected, so that the second factor accounts for most of the residual variance, subject to being uncorrelated with the first factor. This same principle could be applied to selecting additional weights for the additional factors. Principal Components Analysis is recommended when the primary concern is to determine the minimum number of factors that will account for maximum variance in the data for use in subsequent multivariate analysis. The factors are called principal components.

## Interest Rate Risk and PC Components – slope, level and curvature

Interest rate risk is the risk that will worsen the interest-dependent asset, such as loan or bond due to interest rate movements. It also referred to as **market risk**—the longer you hold a bond, the higher increase in market risk.

In general, market interest rates are the most critical factor in the change of bond prices. The change of interest rate leads to the fluctuation of the price of financial products in the fixed-income market. Therefore, it is very important to manage the interest rate risk of the bond portfolio. It can avoid the loss of the portfolio due to interest rate fluctuations. Also, effective estimate can provide a good reference for investment manager in managing assets. An analysis of the changes in the term structure of interest rates is the key part on interest rate risk management.

Traditional interest rate risk management focuses on duration and duration management, which means that only the risk of parallel movement of interest rate maturity structures is considered. However, this management method assumes that only one risk factor affects the change in the term structure of the interest rate, and restricts the term structure of the interest rate only move parallel in one direction, but in fact the interest rate term structure has level and curvature while the risk is being evaluated.

It has been shown that there are three potential uncertainties that influence the intertemporal variation of the interest term structure, which can be interpreted as slope, level and curvature. Many scholars apply principal component analysis (PCA) to the national bond market of different countries to obtain the main factors that affect the interest rate term structure. In general, the interpretation rate of the first three principal components reaches or exceeds 90%.

## Empirical Result

### Application of PCA on Chinese Sovereign Bond Term Structure of Interest Rate

In our research, we use PCA in China Sovereign bond market in order to revealed three principal components (slope, level and curvature) that are sufficient in explaining the variation in the interest rates changes. By using PCA we get a significant reduction in dimensionality when compared with other models. Besides, it is able to produce orthogonal risk factors, this feature makes interest rates risk measurement and management a simpler task, because each rich factor can be treated independently.

**Methods and Steps:**

1. Given the data of China Government Bond Yield rates, and analyze the PCA using the PCA method given by MATLAB

2. In order to calculate Value at Risk of the bond, We need to calculate Principal Component Duration (PCD):

$$\text{PCD}(v) = \sum_{i=1}^{m} \left( KRD_{(i)} \, l_{iv} \right)$$

l = factor loadings (from PCA)

KRD = key rate duration, can be counted using the formula below

$$KRD = \frac{(P(-) - P(+))}{(2 * 1\% * P(0))}$$

P(0) = The original price of the bond

P(-) = The bond price with a 1% decrease in price

P(+) = The bond price with a 1% increase in price

3. Compute of original price of the bond with assumption that all bonds were zero coupon bonds. Face value were assumed to be 1,000.
   With the formula of $P = Face\ value * e^{(-rt)}$, P(0) can be counted by applying the formula with different time to maturity.

4. After PCD for the first three factors are found, Value at Risk can be computed using the formula below.

$$VaR_c = V_0 Z_c \sqrt{\sum_{i=h,s,c} PCD(i)^2}$$

Where
$$Z_c = z\ score\ of\ c\ percentile\ of\ a\ standard\ normal\ distribution$$
$$V_0 = initial\ market\ value\ of\ the\ portofolio$$

**Result:**

- Loadings of the first three principal components of the spot rates:

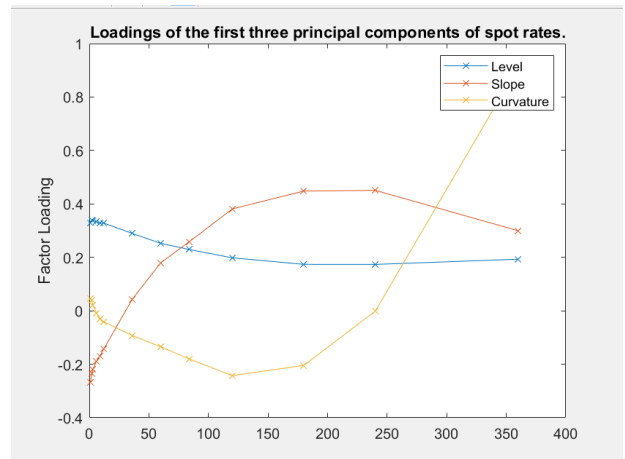

- Percent of var explained by principal component
  Level: 80.8019
  Slope: 14.0525
  Curvature: 2.2670
- Principal Component Duration of the first three factors:
  Level: -0.7344
  Slope: -1.1991
  Curvature: -0.0624
- VaR with 95 percent confidence level is: 231.5430
- VaR with 99 percent confidence level is: 327.3976

## Principle Component-Based Fixed Income Immunization

In many occasion, we need to immunize a liability stream. For this purpose, we create a portfolio of fixed income securities that mimics the liability stream but is exposed less to systematic risks. Then, we hedge our liability stream using the portfolio.

One of the traditional solutions of the problem is to create an immunized portfolio based on Macaulay duration. The problem with using Macaulay duration to create an immunized portfolio is that

Macaulay duration have three assumptions, which are the interest rate's term structure is horizontal and will only move in the vertical direction and the change in interest rate is continuous.

As we can see from the principal component analysis, the first factor can only explain about 60% of the change in the interest rate term structure. This forced us to compare up with a better immunization model to immunize the portfolio.

By introducing the Fisher-Weil duration, the problem can be solved. The Fisher-Weil duration is defined as following:

Assume that the portfolio is consist of N asset or liability that generate cash flow $C_1, C_2, \ldots, C_N$ at time $t_1, t_2, \ldots, t_N$. The corresponding interest rate on time $t_i$ is recorded as $r_0(t_i)$, so the present value of 1 unit of cash at time $t_i$ is $B_i(0) = exp(-r_0(t_i)t_i)$. Then the Fisher-Weil duration is:

$$D_{FW} = \frac{\sum_{i=1}^{N} B_i(0)C_i t_i}{S(0)}$$

Where $S(0)$ is the present value of the portfolio which is $S(0) = \sum_{i=1}^{N} B_i(0)C_i$.

## Multi-factor Immunization Model

From our previous principle component analysis, we have the loading of the first three factor, $U_1, U_2, U_3$ and the three factor, $F_1, F_2, F_3$. Now we are going to propose an immunization model base on the three factor we get.

As we know, the change in the return of each security can be expressed as:

$$X = \begin{pmatrix} X_1 \\ X_2 \\ \vdots \\ X_N \end{pmatrix} = \begin{pmatrix} u_1(t_1) & u_2(t_1) & u_3(t_1) \\ u_1(t_1) & u_2(t_1) & u_3(t_1) \\ \vdots & \vdots & \vdots \\ u_1(t_N) & u_2(t_N) & u_3(t_N) \end{pmatrix} \begin{pmatrix} F_1 \\ F_2 \\ F_3 \end{pmatrix}$$

Where as $U_i = [u_i(t_1), \ldots, u_i(t_N)]'$.

As we know that $U_1, U_2, and U_3$ are orthogonal vectors and any change in the interest rate's term structure could be expressed by a linear combination of the three factors. This make is possible to use each of them to describe a shift in the interest rates' term structure.

We can define a duration by the sensitivity of the portfolio's value to the shift in the interest rate's term structure in different directions. The direction is set by the three factors we got from the principle component analysis. The duration is defined as:

$$D_k^S = \frac{\sqrt{N}}{S(0)} \sum_{i=1}^{N} B_i(0)C_i t_i u_k(t_i)$$

As we can see. This duration defined by the three factor's loading is very similar to the Fisher-Weil duration we introduced before. The only difference is with the factor $u_k(t_i)$. Since $U_1, U_2, and U_3$ are orthogonal vectors, $\sum_{i=1}^{N} u_k^2(t_i) = 1$. So if $u_k(t_i)$ is the same for all k, then $u_k(t_i) = \frac{1}{\sqrt{(N)}}$. Then the duration we define would be the same with Fisher-Weil's duration.

Now our goal is to design a duration immunized portfolio for any kind of cash flow required. In real life application, the portfolio normally is used to match a liability portfolio that will need a stream of future cash flow. We assume this liability portfolio need to pay a cash flow of $C_1^L, C_2^L, \ldots, C_N^L$ at time $t_1^L, t_2^L, \ldots, t_N^L$. Note $S_L(0)$ as the present value of the liability portfolio and $D_k^L$ as its duration at direction $U_k$. The goal is to find an portfolio A such that satisfying the following two conditions:

$$S_L(0) = S_A(0)$$
$$D_K^L = D_k^A \, for \, k = 1,2,3$$

With the result of previous analysis, we know that over 95% of the changes in the interest's term structure can be explained the first three factors, so using $U_1, U_2, U_3$ can give a pretty good immunized portfolio already. If higher precision is required, then this model is open to add more factors in to increase explanation power and decrease risks.

## Conclusion

Principal Component Analysis has been widely used to study the shift in the term structure of interest rate. We have used PCA to identify the factors which are responsible for changes in yield curve. The results indicate that the three factors provide us the most of the variations in the term structure shift in Chinese Sovereign Bond Market. The study finds that the first three principal components explain a major part of the total variance of interest rate changes. This result is consistent with other studies. The first factor accounts for 80.80% of the total variance, while the second and third factors account for 14.05% and 2.27%, respectively. In sum, the first three principal components explain 97.12% of the variability of the data, which indicates that these factors are sufficient for describing the changes in the interest rate term structure in Chinese Sovereign Bond Market.

## Appendix

Matlab Code

```matlab
% Computer Programming in Financial Engieering
%
%=================================================

clear
close all

%=================================================
%
% The Excel file with spot rates.  This points to the directory
% where I keep it on my computer.  This location needs to be
% changed to whatever location you use.
%
%=================================================

dataFile = 'china_gov_bond_yield.xlsx';

%=================================================
%
% Read the spot rates, their maturities, and the dates
%
%=================================================

maturitiesInMonths = xlsread(dataFile, 'B1:N1');
spotRates = xlsread(dataFile, 'B4:N4248');
```

```matlab
%getting the dates from excel to number(in Raw)
[Sig, TStr, Raw] = xlsread(dataFile,'A4:A4248');

matlabDateNumber = datenum(Raw);
matlabDataStr=datestr(matlabDateNumber);

%%

%================================================================
%
% Sample means and standard deviations
%
%================================================================
%%
[numObs, numBonds] = size(spotRates); %numObs is row (195) numBonds is column (30)

meanYields = mean(spotRates);
stdYields = std(spotRates);

monthlyRateChanges = spotRates(2:end,:) - spotRates(1:(end-1),:);

%%
%================================================================
%
% Contemporaneous correlations in levels and first differences
%
%================================================================

corrMatrix = corr(spotRates);
diffCorrMatrix = corr(monthlyRateChanges);

%%
% Using the full sample of data, construct a figure that displays loadings
% of the first three principal components of spot rates

%================================================================
%
% Principal components
%
%================================================================

[coefMatrix, score, latent, tsquared, explainedVar] = pca(spotRates);

factors = spotRates * coefMatrix;

[U,E] = eig(cov(spotRates));

figure(3)

plot(maturitiesInMonths, coefMatrix(:,1:3), '-x');
ylabel('Factor Loading');
```

```matlab
legend('Level', 'Slope', 'Curvature');
title('Loadings of the first three principal components of spot rates.')

print('prinCompFig', '-djpeg');
print('prinCompFig', '-depsc');

%%
% Construct two tables that contain information about the relative
importance
% of the first three factors (percent of total variance explained)
% and the magnitude of unexplained variation in monthly changes spot
% rates (standard deviation of residuals). You only need to write down
% the table for spot rates with maturities up to 10 years

% amount of variation explained by each PC.

fprintf(1, 'Percent of var explained by principal component\n');
disp(explainedVar(1:3)) %Only want the first three factor!

%% RESIDUAL PART
%=========================================================================
% The principal components are treated as factors
% Now use factors found by pca to explain variations in interest rates
% amount of variation in yields not explained by the first 3 level
% factors.  Square root of estimated variance from regression
% (corrects for degrees of freedom, but not construction of factors)
%
%=========================================================================

residSD = zeros(numBonds, 1);
gamma1=zeros(numBonds,1);
gamma2=zeros(numBonds,1);
gamma3=zeros(numBonds,1);

constant = ones(numObs,1);
for i = 1:numBonds
   [b, bint, r, rint, stats] = ...
       regress(spotRates(:,i), [constant, factors(:,1:3)]);
   residSD(i) = sqrt(stats(4));
   gamma1(i)=b(2);
   gamma2(i)=b(3);
   gamma3(i)=b(4);
end

fprintf(1, '\n\n');
fprintf(1, 'Standard deviations of yields of bonds (basis points) \n');
fprintf(1, ...
        '  Maturity   SD    SD of resid from 3-factor\n');
fprintf(1, ...
        '-----------------------------------------------------\n');
for i = 1:10 %Only want first 10 years
   fprintf(1, ' %4d     %5.0f      %6.1f \n', ...
```

```matlab
            maturitiesInMonths(i), 10000*stdYields(i), ...
            10000*residSD(i));
end

%% Principal Component Duration

principal_component_duration=zeros(3,1);
for i = 1 : 3 % there are 3 factors
    sum_temp = 0;
    for j = 1 : 13 %Count sigma
        face_value = 1000; % assume to be 1000 with zero coupon bonds
        rt = 0.55;%discount rate (whatever number I put does not play significant role)
        tm = maturitiesInMonths(j); %time to maturities
        p_zero = face_value * (double(eulergamma).^(-rt*tm));
        p_minus = 0.99*p_zero;
        p_plus = 1.01 * p_zero;

        % Key rate duration = (P(-) - P(+)) / (2 x 1% x P(0))
        key_rate_duration = (p_minus - p_plus) / (2 * 0.01 * p_zero);
        % factor loading is in coef matrix
        key_multiplied_factors_loadings = key_rate_duration*coefMatrix(i,j);
        sum_temp = sum_temp +key_multiplied_factors_loadings;

    end
    principal_component_duration(i) = sum_temp;

end

%% Value at Risk
% Value at Risk count

% Assume initial market value of the portofolio is 1,000
% VaR at 95% confidence interval
VaR_95 = 1000 * 1.645 * sqrt(principal_component_duration(1).^2 + principal_component_duration(2).^2 + principal_component_duration(3).^2)*0.1;
fprintf("\n\nVaR with 95 percent confidence level is : %8.4f \n",VaR_95 );

% VaR at 99% confidence interval
VaR_99 = 1000 * 2.326 * sqrt(principal_component_duration(1).^2 + principal_component_duration(2).^2 + principal_component_duration(3).^2)*0.1;
fprintf("VaR with 99 percent confidence level is : %8.4f \n",VaR_99);
```